\documentstyle[psfig,preprint,aps,prb]{revtex}
\draft

\begin{document}

\title{Phase diagram and hidden order for generalized spin ladders}

\author{S. Brehmer, H.-J. Mikeska, U. Neugebauer}
\address{Institut~f\"ur~Theoretische~Physik, Universit\"at~Hannover, 
30167~Hannover, Germany}

\maketitle

\begin{abstract}
        We investigate the phase diagram of antiferromagnetic 
        spin ladders with additional exchange interactions on 
        diagonal bonds by variational and numerical methods. 
        These generalized spin ladders interpolate smoothly 
        between the $S=1/2$ chain with competing nn and nnn 
        interactions, the $S=1/2$ chain with alternating exchange
        and the antiferromagnetic $S=1$ chain. 
        The Majumdar-Ghosh ground states are formulated 
        as matrix product states and are shown to exhibit the same type of 
        hidden order as the af $S=1$ chain. Generalized matrix product
        states are used for a variational calculation of the ground
        state energy and the spin and string correlation functions.
        Numerical (Lanczos) calculations of the energies of the ground 
        state and of the low-lying excited states are performed, and 
        compare reasonably with the variational approach. 
        Our results support the hypothesis that the dimer and 
        Majumdar-Ghosh points are in the same phase as the af $S=1$ chain.
\end{abstract}

\medskip


\section{Introduction}
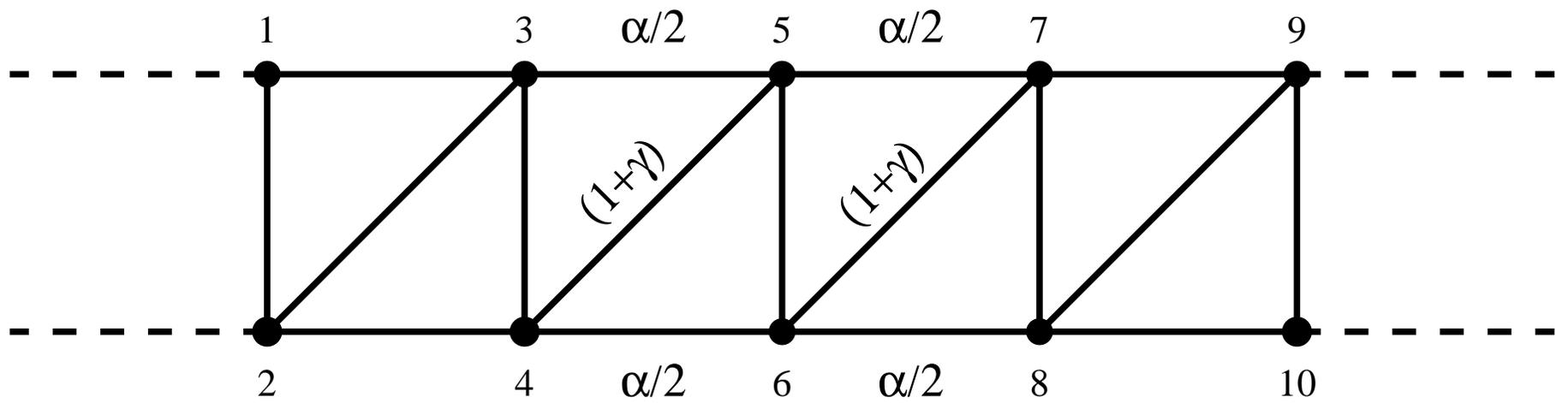
\begin{figure}
\caption{Structure of the  generalized spin ladder with additional 
        diagonal interaction.
\label{ladder}}
\end{figure}

In the last few years spin systems consisting of a finite 
number $n$ of interacting spin chains with $S=\frac{1}{2}$ 
(now usually called spin ladders, consisting of legs and rungs, see 
fig.~\ref{ladder} for $n=2$) have attracted considerable attention 
as recently reviewed by Dagotto and Rice \cite{DagR96}.
>From the theoretical point of view, these systems are
intermediate between one- and two-dimensional systems. 
Interest in these systems started with the two leg ladder \cite{Hir88}
with isotropic antiferromagnetic couplings 
on the legs and rungs; in contrast to the $S=\frac{1}{2}-$chain this 
ladder is characterized by a spin gap. The appearance of this gap is 
immediately clear in the limit of strong ferromagnetic interactions 
on the rungs (it then is identical to the gap characterizing the
$S=1$ antiferromagnetic (Haldane) chain) and in the limit of 
interactions on the rungs only (dimer limit); the gap, however, 
apparently persists not only for the isotropic antiferromagnetic ladder 
\cite{GopRS94,Hid95,Whi96} but for 
nearly all values of the coupling constants except in the limiting case
of two decoupled chains. In more recent investigations, af ladders with 
$n$ legs have been shown to behave as spin liquids for $n$ odd and
as gapped systems for $n$ even \cite{FriAT96,HatN95}.

Experimental investigations have been performed on the $n=2$ ladder
substance $\rm (VO)_2P_2O_7$ \cite{JohJGJ87} and also on the family of
compounds $\rm Sr_{m-1}Cu_{m+1}O_{2m}$ ($m=3,5,7...$) \cite{HirATB91}
which approximately realize ladders with $n= \frac{1}{2}(m+1)$
legs. In these materials evidence for the existence of a spin gap has
been found in susceptibility \cite{JohJGJ87,HirATB91}, neutron scattering
\cite{EccBBJ94} and magnetic resonance 
\cite{Kat95,Lut96} experiments. The cuprates are of additional interest as
candidates for high temperature superconductivity, which, however, remains
undetected \cite{HirT95}.

These experimental results have been successfully dealt with in 
a large number of theoretical approaches using e.g. perturbation 
theory and exact numerical 
diagonalization \cite{BarDRS93,BarR94}, mean field theory based on 
bond operators \cite{GopRS94} and the density matrix 
renormalization group approach \cite{Hid95,Whi96}.  

Recently theoretical interest in spin ladder models with $n=2$ has
concentrated on the fact that these models allow to interpolate
smoothly between seemingly very different spin systems when the rung
coupling is varied, such as between the $S=1$ antiferromagnetic
(Haldane) chain and the dimer state, i.e. models which are built of
triplets, resp.\ singlets on the rungs \cite{Hid95,Whi96,Wat96}.

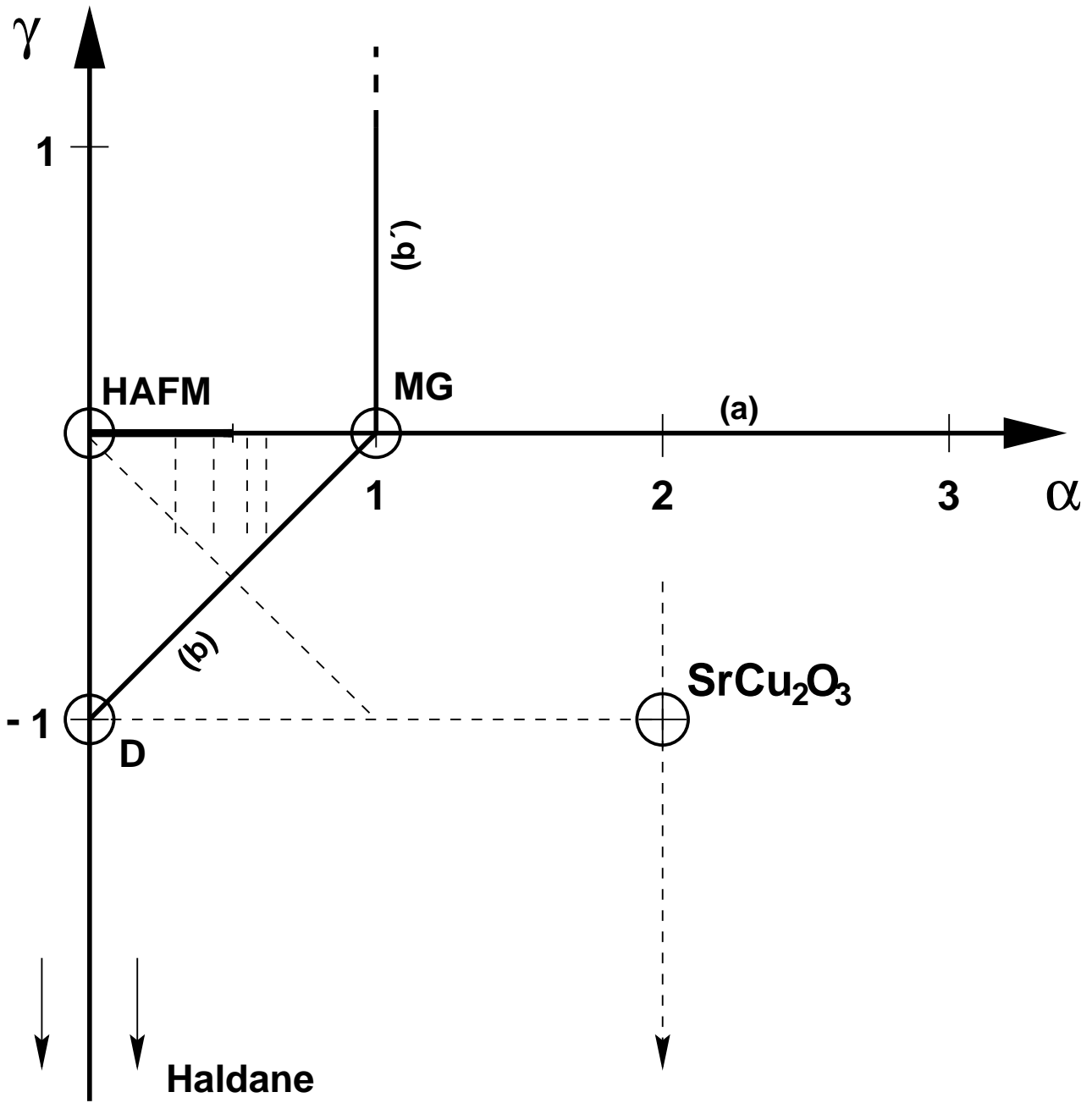
\begin{figure}
\caption{The phase diagram of the generalized spin ladder. Paths where 
        numerical calculations have been performed are shown as dashed lines.
\label{phasediag}}
\end{figure}

An interpolation between these two limiting 
models which avoids the singular point of two decoupled legs
is possible when an additional interaction on one of the diagonal 
bonds is introduced as proposed by
White \cite{Whi96} and Chitra et al \cite{ChiPKSR95}. This model is 
defined by the following Hamiltonian (see fig.~\ref{ladder}):

\begin{equation}
        H=H^{(0)}+ \frac{\alpha}{2} H^{(1)}+ (1+\gamma) H^{(2)}
\label{hami} 
\end{equation}
\begin{equation}
        H^{(0)}= \sum_{j=1}^L \vec{S}_{1,j} \; \vec{S}_{2,j} \qquad
        H^{(1)}= \sum_{j=1}^L \sum_{i=1}^2 \vec{S}_{i,j} \; \vec{S}_{i,j+1}
        \qquad H^{(2)}= \sum_{j=1}^L \vec{S}_{1,j} \; \vec{S}_{2,j+1}
\end{equation}
Periodic boundary conditions are used and operators $\vec{S}_{i,j}$ 
denote spin--1/2 operators with the
indices $i = 1,2$ labeling the two legs and $j=1,2 \dots L$ labeling
the $L$ rungs. When sites are labeled
by a single index (running from 1 to $2L$ as also indicated in
fig.~\ref{ladder}) it becomes clear that the spin ladder with 
additional diagonal couplings is a generalized model composed of 
the spin--1/2 chains with bond alternation and next nearest
neighbour (nnn) interaction. This model includes in particular 
(compare with fig.~\ref{phasediag}):

\begin{itemize} 
\item the antiferromagnetic (af) $S=\frac{1}{2}$ Heisenberg 
chain with nnn exchange of relative strength $\frac{\alpha}{2}$ 
($\alpha$-axis, $\gamma = 0$)), including in particular the 
Majumdar-Ghosh (MG) limit \cite{MajG69} 
($\alpha= 1, \gamma = 0$) with two degenerate dimerized
exact groundstates given as a product of singlets 
either on the rungs or on the diagonals,

\item the af $S=\frac{1}{2}$ Heisenberg chain with alternating exchange 
($\gamma$--axis, $\alpha = 0$), including in particular the 
dimerized chain ($\alpha = 0, \gamma = -1$) with couplings 
on the rungs only and a product of singlets on the rungs
as exact ground state,  

\item the ''regular'' isotropic $S=\frac{1}{2}$ ladder for $\gamma=-1$; 
the experimentally relevant cases
mentioned above are expected to correspond to $\alpha = 2$, i.e.to 
coupling of equal strength on the rungs and on the legs,  

\item the Haldane chain with $L$ sites for $\gamma \to -\infty$,
$\alpha > 0$ and nondiverging (leading to $S=1$ units with an
effective af coupling $\frac{1}{4} (1+\alpha)$ on the diagonals),

\item the isotropic $S={1 \over 2}$ af Heisenberg chain
with $2L$ sites (for $\alpha = 0, \; \gamma = 0$) and two decoupled
af Heisenberg chains with $L$ sites each (for $\alpha \to \infty$ with
$\gamma$ finite).
\end{itemize}

In addition the following lines in the phase diagram are of 
particular interest:

\begin{itemize}

\item the line (a), $\gamma = 0,\: \alpha \ge 0$, is of particular interest
since on this line a transition from the gapless isotropic af Heisenberg 
chain to the gapped MG chain occurs with increasing nnn
coupling strength $\alpha$. This transition has been
investigated in detail \cite{Hal82,TonH87} (for a review see 
ref.\onlinecite{TonH92}) and the critical point has
recently been located very accurately at 
$\frac{1}{2} \alpha_{cr} \approx 0.241167$ \cite{Egg96}. 
Line (a) and its immediate neighbourhood ($\gamma \ll 1$) is also of 
experimental relevance because of the discovery of the spin-Peierls compound 
$\rm CuGeO_3$ with alternating and strong nnn interactions \cite{CasCE95};

\item the straight line (b), which connects the dimer point to the
MG point ($(\alpha, \gamma) = (0,-1) \to (1,0)$); on this line one of 
the MG ground states, the state with singlets on the rungs
remains an exact ground state \cite{ShaS81}. 
\end{itemize}

In this paper we will discuss the properties of the generalized
spin ladder in the space spanned by the interaction constants 
$\alpha > 0$ and $\gamma$.
First we note that there is a symmetry transformation connecting 
the upper half-plane of the phase space of fig.~\ref{phasediag} 
to the strip $-1 < \gamma < 0$: a translation of the 
upper leg by one lattice constant to the left leads to a ladder with 
the same symmetry, but different coupling constants. The symmetry 
transformation exchanges the roles of rungs and diagonals; apart
from a change in energy scale by a factor $1 + \gamma$ (which is 
qualitatively irrelevant for $\gamma > -1$), it is expressed in the
following way in terms of the parameters of the Hamiltonian: 

\begin{equation}
\gamma \to -\frac{\gamma}{1+\gamma}, \qquad \alpha \to 
                  \frac{\alpha}{1+\gamma}.   \label{transf} 
\end{equation}

The symmetry line of this transformation 
is the $\alpha-$axis $\gamma = 0$, points on this axis are mapped 
onto themselves. Owing to the presence of 
this symmetry we will consider in the following the region $\gamma < 0$
of the $\alpha-\gamma-$plane only. The strip $-1 < \gamma < 0$ is 
related to the upper half-plane by the symmetry of eq.(\ref{transf}),
the section $\gamma > -1$ of the $\alpha-\gamma$--plane has af coupling
and is identical to the phase space discussed by Shastry and Sutherland
\cite{ShaS81}. Including the region $\gamma < -1$ adds the possibility of 
ferromagnetic coupling between the two legs, this possibility has so far 
been discussed for special cases only \cite{Wat94,ChiPKSR95}.
  

In the following we will present results for the generalized spin
ladder as defined by the Hamiltonian of eq.(\ref{hami})
from a variational approach based on the concept of hidden
order. These results will be supplemented by and compared to the 
results from exact diagonalizations for finite ladders with up to
14 rungs. The question of hidden order in 
generalized spin ladders comes up naturally since
the af $S=1$ chain, the prototype system where hidden order has emerged
first, is realized in the limit $\gamma \rightarrow - \infty$. 
In the Haldane chain \cite{Hal83}
hidden order has been made transparent by Kennedy and Tasaki
\cite{KenT92}, using a nonlocal unitary transformation. Ground states
with complete hidden order were then written down in the form of
matrix product (MP) ground states \cite{KluSZ93} and used as
basis for variational calculations. The new feature of
ladders is that singlets on the rungs get mixed in when one moves from
the Haldane limit $\gamma \rightarrow - \infty$ towards finite values
of $\gamma$. A generalization of the Kennedy-Tasaki-transformation
\cite{KenT92} was proposed by Takada and Watanabe \cite{TakW92} for
the isotropic spin ladder ($\gamma = -1$, varying $\alpha$).
Explicit definitions of hidden order in spin ladders were recently
discussed by Nishiyama et al \cite{NisHS95} 
and White \cite{Whi96}. 

Our variational approach starts from the observation that the two
exact ground states of the Majumdar-Ghosh Hamiltonian can be written 
in the form of matrix product states. We then generalize the matrix 
product state to allow the ansatz to include the approximate ground 
state of the Haldane chain
as given by Affleck et al \cite{AffKLT88} and written as matrix 
product state by Kl\"umper et al \cite{KluSZ93}. Continuous variation 
of the parameters then smoothly connects the seemingly unrelated limiting
cases of the Majumdar-Ghosh and dimer points on the one hand and of the
Haldane chain on the other hand.
We expect this ansatz to make sense as a frame  
for a qualitative description in a large part of the half plane
$\alpha > 0$. 


In section 2 we will introduce the variational matrix product states
defining the hidden order in the analogous way as for an S=1 chain but
replacing the contribution of the $S^z=0$ component of the rung
triplet state by a linear combination of this component with the rung
singlet. The exactly known ground states of the Majumdar-Ghosh chain
are particular examples of this ansatz. This definition of matrix product
states imposes a particular type of hidden order which we will
describe. In section 3 we present the results of variational
calculations for the ground state energy and for ground state
correlation functions obtained by using this scheme. In addition the
string correlation function as proposed in ref.\onlinecite{Whi96} will
be computed.  In section 4 we present numerical results for the ground
state energy and the gap along various paths in the phase diagram
(shown as dashed lines in fig.~\ref{phasediag}). These results are
obtained from exact diagonalization of finite ladders with up to 14
rungs and periodic boundary conditions using the Lanczos algorithm;
they will be discussed in comparison to the results of the variational
calculations. Our results imply that a smooth variation of parameters
leads one from the Haldane-phase to the dimer-phase; they therefore
support the hypothesis that these two limiting models are not
separated by a phase transition and that the only critical behaviour
in the phase diagram considered is on the gapless line $\gamma = 0, \;
0 < \alpha < \alpha_{cr}$. Special attention will be paid to the
neighborhood of this line in the numerical calculations. Conclusions
will be given in section 5.\\

\section{Matrix-Product States for Spin Ladders}

We start by a discussion of the Majumdar-Ghosh chain, 
i.e. $\alpha = 1, \gamma =0$.
For periodic boundary conditions the two ground states are known exactly:
they are dimerized configurations with singlets on either the
rungs, $|0_r \rangle$, or the diagonal bonds, $|0_d \rangle$. 
If we use the following representation of states
on the j-th rung,

\begin{eqnarray}
       |s \rangle_j &=&  \frac{1}{\sqrt{2}} ( | \uparrow \downarrow \rangle_j
       - | \downarrow \uparrow \rangle_j ), \nonumber\\
       |t_0 \rangle_j &=&  \frac{1}{\sqrt{2}} ( | \uparrow \downarrow \rangle_j
       + | \downarrow \uparrow \rangle_j ), \qquad              
       |t_+ \rangle_j =  | \uparrow \uparrow \rangle_j, \qquad 
       |t_- \rangle_j =  | \downarrow \downarrow \rangle_j,
                                                    \label{representation}
\end{eqnarray}
the two ground states can be writen in form of MP-states

\begin{equation}
        |0_r \rangle = \prod_{j=1}^L |s \rangle_j \qquad
        |0_d \rangle = (-1)^L \;  Tr\left( \prod_{j=1}^L g_j \right)
                                             \label{orderst}
\end{equation}
with the matrices $g_j$ defined by

\begin{equation} 
        g_j = \frac{1}{2} \left( \begin{array}{cc}
        |s \rangle_j + |t_0 \rangle_j & - \sqrt{2} |t_+ \rangle_j \\
        \sqrt{2} |t_- \rangle_j & |s \rangle_j - |t_0 \rangle_j
        \end{array} \right).
                                            \label{mgmatrix}
\end{equation}

In the representation of eq.(\ref{orderst}) both MG ground states are
given in terms of singlets resp.\ triplets on the {\em rungs}; this 
appears complicated for the state $\vert 0_d \rangle$ but
actually it serves to realize the presence of hidden order in the MG
ground states: The matrix representation of the ground state according to
eqs.(\ref{orderst},\ref{mgmatrix}) has components of the following
string structure only   

\begin{eqnarray}
\scriptstyle{
        \ldots t_+ \; t_- \; t_+ \; t_- \; (t_0+s)(t_0+s)  \ldots \ldots  
        (t_0+s)(t_0+s) \; t_+ \; t_- \; t_+ \; t_- \; t_+ \;
        (t_0-s)(t_0-s)  \ldots \ldots 
        (t_0-s)(t_0-s) \; t_- \; t_+ \; t_- \; t_+ \ldots ,
}
\end{eqnarray} 
i.e. the sequence $\ldots t_+ \; t_- \; t_+ \; t_- \; \ldots $ is
interrupted either by an arbitrary number of sites with $\vert t_0+s
\rangle$ following a site with $\vert t_- \rangle$ or by an arbitrary
number of sites with $\vert t_0-s \rangle$ following a site with
$\vert t_+ \rangle$. 

The symmetry transformation of eq.(\ref{transf}) transforms the ladder 
into itself on the line $\gamma = 0$ and the two states $\vert
0_s \rangle$ and $\vert 0_d \rangle$ remain degenerate 
for $\alpha \ne 1$; they are,
however, ground states only at the Majumdar-Ghosh point. When 
moving off the axis $\gamma=0$ this symmetry is
destroyed and at most one of these states survives as ground state.
In particular, on the line (b), $(\alpha, \gamma) = (0,-1) \to (1,0)$ the
ground state is $\vert 0_r \rangle$, i.e.
composed of rung singlets. This 'disorder line' is therefore
characterized by a vanishing of the correlation length.  
Under the symmetry transformation (\ref{transf}) 
this line (b) transforms into the line (b')
$(\alpha, \gamma) = (1,0) \to (1,\infty$) with a dimerized ground state
with singlets located on diagonal bonds. 
As discussed in the introduction it is sufficient
to discuss only the region $\gamma < 0$ in the phase diagram of
fig.~\ref{phasediag}.

When we move away from the MG point, the state 
$\vert 0_r \rangle$ has no freedom to adapt but the state
$\vert 0_d \rangle$ can be used as the basis for a variational calculation
by allowing different amplitudes for the singlet and the triplet
contributions. Rotational invariance requires the following form for the
matrix $g_j$ 

\begin{equation}
        |\psi \rangle = Tr \left( \prod_{j=1}^L g_j \right),
        \qquad g_j = \left( \begin{array}{cc}
        a|t_0 \rangle_j + b|s \rangle_j & -a \sqrt{2} |t_+ \rangle_j \\
        a \sqrt{2} |t_- \rangle_j & -(a|t_0 \rangle_j- b|s \rangle_j)
        \end{array} \right)                    \label{wavefunction}
\end{equation}

This follows from the fact that the three states 

\begin{eqnarray}
 \vert t^{x} \rangle &=& -\frac{1}{\sqrt{2}} ( \vert \uparrow \uparrow \rangle
        - \vert \downarrow \downarrow \rangle ),     \qquad 
 \vert t^{y} \rangle = \frac{i}{\sqrt{2}} ( \vert \uparrow \uparrow \rangle
        + \vert \downarrow \downarrow \rangle ), \nonumber \\
 \vert t^{z} \rangle &=& \frac{1}{\sqrt{2}} ( \vert \uparrow \downarrow 
        \rangle + \vert \downarrow \uparrow \rangle ) =: \vert t_{0} \rangle
\end{eqnarray}
form a vector and can only appear in the $2 \times 2$ matrix $g_j$ in
the combination $\sum_{\alpha} \sigma^{\alpha} \vert t^{\alpha} \rangle $ 
in the case of rotational invariance ($\sigma^{\alpha}$ are the Pauli 
spin matrices). In eq.(\ref{wavefunction}), $a$ and $b$ are complex 
variational parameters, subject to the normalization
condition $3 \vert a \vert^2 + \vert b \vert^2 = 1$. Thus the parameters
entering the variational calculation are one amplitude and one phase.

The ansatz of eq.(\ref{wavefunction}) can be shown to reduce to the state
introduced by Takada and Watanabe \cite{TakW92} when their state 
is written in
original spin space (no unitary transformation) and perfect generalized
hidden order is introduced. If we put $b=0$ in 
eq.(\ref{wavefunction}) we obtain the AKLT state \cite{AffKLT88}, 
the exact ground state of the af $S=1$ chain with biquadratic exchange 
of strength $\frac{1}{3}$.

It has to be noticed that the variational ansatz of
eq.(\ref{wavefunction}) may be used with the matrices $g_j$ defined 
either on the rungs or on the
diagonals and as a first step of the variational calculation 
it is necessary to find out which of these possibilities
leads to lower energy. It turns out that for $\gamma < 0$ 
(which is the region we will consider)
lower energy is obtained when $g_j$ is defined on the 
diagonals, whereas defining $g_j$ on rungs gives the lower energy
for $\gamma > 0$. On the symmetry line $\gamma = 0$
eq.(\ref{wavefunction}) is an approximate description of the two
equivalent groundstates for $\alpha_{cr} < \alpha \ll \infty$.
The ansatz of eq.(\ref{wavefunction}) allows to reproduce the exact
ground state on line (b) ($\gamma = -1 + \alpha$), the variational 
approach is therefore exact on this line.    

On the one hand, the wave function of eq.(\ref{wavefunction}) is
rather general since it can smoothly interpolate between dimerized 
states and AKLT-states, i.e.\ states which correspond to points 
which are far separated in the phase diagram. On the other hand 
one cannot expect this approach
to work for the whole $\alpha$--$\gamma$--plane since it is manifestly
inadequate for special points like the ones corresponding to
the HAF or to two decoupled chains, which have gapless spectra 
and power-law decay of correlations functions. 

We finally note that there is another possibility of constructing 
a rotationally invariant wave function of the MP type which is related to
the resonating valence bond structure also discussed for spin ladders:
a MP wavefunction with the following alternating structure

\begin{equation} 
\vert 0_{RVB} \rangle = Tr \left( \prod_{k=1}^{\frac{L}{2}} g^{+}_{2k-1} 
      g^{-}_{2k} \right), \; \; \;
      g^{\pm}_j = \frac{1}{2} \left( \begin{array}{cc}
        \pm |s \rangle_j + |t_0 \rangle_j & - \sqrt{2} |t_+ \rangle_j \\
        \sqrt{2} |t_- \rangle_j & \pm |s \rangle_j - |t_0 \rangle_j
        \end{array} \right),             
        \label{alternate} 
\end{equation}
i.e.\ the previous structure with diagonal elements interchanged in
every second g-matrix, describes a state with singlets located on 
the legs. Thus our ansatz is also connected to RVB states on ladders 
as introduced in refs.\onlinecite{WhiNS94,ZenP95}. 

\section{variational calculations for generalized spin ladders}  

In this section we present the calculation of the energy, of the spin
correlation function and of the string (hidden order) correlation function
using the variational state of eq.(\ref{wavefunction}). We discuss
results for these quantities in the $\alpha-\gamma$--plane and compare
to the results of alternative approximate approaches for special points. 
A comparison to the results of numerical calculations will be
given in section 4. 
 
The calculation of
the ground state energy and of the ground state correlation functions
can be done in complete analogy to the corresponding calculations
for the $S=1$ af chain \cite{KluSZ93}. 
The variational energy, i.e.\ the expectation value 
of the Hamiltonian of eq.(\ref{hami}) with the wavefunction of 
eq.(\ref{wavefunction}) and the matrix $g_j$ defined on diagonal 
bonds, is obtained as

\begin{eqnarray}
       \frac{E_{var}}{L} & = & -3 \{A^4 + 2 A^3 B \cos \varphi 
                  + A^2 B^2 \cos^2 \varphi \} \nonumber \\
       &   &  -3 \alpha \{A^4 - A^2 B^2 \cos^2 \varphi \}       
              + \frac{3}{4} (1+\gamma) \{A^2 - B^2\}, \\
       &   &  A = \vert a \vert, \:  B = \vert b \vert, \nonumber
         \label{varE}
\end{eqnarray}
with the norm $3 A^2 + B^2 = 1$ as an additional constraint.  
$\varphi$ is the relative phase of the original complex amplitudes 
$a$ and $b$. 

$E_{var}$ reduces to the expression given by Watanabe \cite{Wat96} for
$\gamma=-1$ and to the expression given by Takada and Watanabe
\cite{TakW92} for $\gamma \to -\infty, \; \alpha \propto (1 +
\gamma)$ for perfect generalized hidden order. The procedure of
minimizing $E_{var}$ is performed easiest by first minimizing with
respect to $\cos \varphi$ and then with respect to the ratio
$u=B/A$. The analysis shows that the absolute minimum always
corresponds to $\cos \varphi = \pm 1$, and the amplitudes $a$ and $b$
can always be taken real; the remaining equation for $u$ is of third
order. Thus the minimum can be calculated exactly for general values
of $\alpha$ and $\gamma$.

The simple variational wave function of eq.(\ref{wavefunction})
also allows the calculation of the ground state correlation functions.
They exhibit the exponential decay
characteristic for MP states. Owing to isotropy there is only one 
correlation length which is obtained as 

\begin{equation}
  \xi_{long}^{-1} = \xi_{trans}^{-1} = 
                   \ln \left( \frac{1}{\vert 1 - 4 A^2 \vert} \right) 
                               \label{lcorrelation} 
\end{equation}

In order to discuss the hidden order for the ladder system, we use the
definition given in ref.\onlinecite{Whi96}, which reproduces the
known results in the $S=1-$chain limit. Using this definition, we
obtain the following result for the string correlation function 
(in the thermodynamic limit $L \to \infty$)

\begin{equation}
        |g(l)| := | \langle (S_{2,0}^z+S_{1,1}^z) 
                  e^{i \pi \sum_{k=1}^{l-1}
                (S_{2,k}^z+S_{1,k+1}^z)} (S_{2,l}^z+S_{1,l+1}^z)
                \rangle | = \frac{4}{9} (1-B^2)^2. 
                \label{string} 
\end{equation}

In fig.~\ref{symmetryline} we show results obtained from the MP state 
ansatz for points on the lines $\gamma = 0, -1$.
In the MG-case ($\gamma = 0, \alpha =1$) and at the dimer point the
variational ground state is exact, we have $A = B = \frac{1}{2},
\; \cos \varphi = 1$. The wavefunction is identical to that of 
eqs.(\ref{orderst},\ref{mgmatrix}), one of the two exact MG ground state
wavefunctions with energy $E_{MG} = - \frac{3}{4} L$. The correlation 
lengths $\xi_{MG} = \xi_{dimer}$ vanish. 
Figs.~\ref{symmetryline}(a,b) show the ground state energy and the 
correlation length on the lines $\gamma=0, -1$. 
Discussing our results on the line $\gamma = 0$ i.e.\ for the general 
af chain with nnn interactions, we have to restrict the discussion 
to the range $\alpha > \alpha_{cr} \approx 0.5$ since the MP ansatz
is no more appropriate when the gapless system is approached. 
Nevertheless we note that the variational ground state energy
for $\alpha = \alpha_{cr}$ is $E_0/L = -0.79354$, i.e.\ within 1\% of the
exact value \cite{ChiPKSR95}. In the region to the right of the disorder
line the relative error increases to typically 5\%. The correlation 
lengths remain very small as is typical for MP wave functions.

The string order parameter is determined by the singlet weight $B^2$ 
which is shown in fig.\ref{symmetryline}c, we have 
$B^2 \ge \frac{1}{4}$ to the left and 
$B^2 \le \frac{1}{4}$ to the right of the disorder line (b).
The string order parameter for the two degenerate MG states takes the
values $g_{0_r}(l)=1/4, \; g_{0_d}(l)=0$ (obtained from 
$B_{0_r}=1/2, \; B_{0_d}=1$). The fact that the string correlation
is different for the two equivalent ground states is related to the
asymmetric definition of the string correlation with respect to rungs
and diagonals in eq.(\ref{string}).

For the symmetry line $\gamma = 0$, another variational wavefunction
of RVB type (which is also exact at the MG point) was proposed by Zen
and Parkinson \cite{ZenP95}. When the results for the ground state
energy are compared for these two approaches, for $\alpha > 1$ lower
energies are found from the RVB approach and for $\alpha < 1$ from the
present MP state approach. The correlation length for $\alpha > 1$ is
similar in the RVB and the MP state approaches, wheras for $\alpha <
1$ the MP ansatz results in a larger increase in $\xi$.  The behaviour
of the correlation lengths in this region of the phase diagram might
be influenced by the possible emergence of a spatially modulated
ground state (spiral phase) at $\alpha=1$ \cite{ChiPKSR95,ZenP95}. The
MP state ansatz cannot by construction reproduce such a behaviour and
the alternating ansatz as described above in eq.(\ref{alternate}) does
not lead to lower variational energies either (except for large values
of $\alpha$). A more detailed discussion of this aspect will be given
when we discuss our numerical results in section 4.

We now turn to a discussion of results for the regular ladder system
$(\gamma=-1,\; \alpha=2)$. From the variational approach we obtain 
$E_0/L \approx -1.102$ and $\xi \approx 0.815$ for this isotropic ladder.
These results may be compared to those for a variational RVB wave 
function which was proposed by Fan and Ma \cite{FanM88} and leads to 
a ground state energy of $E_0/L \approx -1.112$ and a correlation 
length of $\xi \approx 0.238$.  The difference between the two
variational energies of about 1\% appears negligible when it is
compared to the exact (numerical) result $E_0/L \approx -1.156$. Both
variational methods lead to correlation lengths considerably
smaller than $\xi \approx 3.19$ as computed in
ref.\onlinecite{WhiNS94}. This is consistent with the well-known fact
that in approaches using MP states the correlation length is notoriously 
underestimated (compare the AKLT value of $\xi \approx 0.9102$ 
to the exact correlation length $\xi \approx 6.2$ in the af $S=1$ chain).


Finally we want to illustrate in the set of figs.~\ref{figsing}(a-c)
how the limit of an antiferromagnetic $S=1$
chain is included in the present scheme for $\gamma \rightarrow
-\infty$. In this limit it is very unfavorable to have singlets on
the diagonal bonds, so minimization leads to $B \rightarrow 0$. 
We show the ground state energy and the singlet weight  
with increasing $\vert \gamma \vert$ for $\alpha = 2$
in fig.~\ref{figsing}(a) and in fig.~\ref{figsing}(c) respectively. Combining
these graphs with the corresponding ones from fig.~\ref{symmetryline}
we can follow a continuous path from the dimer point to the Haldane 
limit. The smooth variation of the physical quantities on such a path 
illustrates the similarity of the Haldane phase and the dimer 
phase in the present approach. 

The minimum of the variational energy in the limit appropriate for the af
$S=1-$ chain has the following form 

\begin{eqnarray}
   E_{0}^{S=1}&=& \lim_{\gamma \to -\infty} E_{0} \approx 
                \left( \frac{1+\gamma}{4} - \frac{\alpha+1}{3}
        + \frac{1}{3 \; (1+\gamma)} \right) \: L 
          \qquad \quad (\vert 1+\gamma \vert \gg 1) \nonumber \\
   B^2 & \approx & \frac{1}{3 \; (1+\gamma)^2}
\end{eqnarray}

The first term in $E_0^{S=1}$ is just the internal energy of the
triplets representing the internal energy of spins $S=1$ at the $L$
sites. We see that minimization in the limit of infinite negative 
$\gamma$ leads to the AKLT-state with

\begin{equation}
A^2=\frac{1}{3}, \qquad E_0 = -\frac{4}{3} J_{eff}, \; \qquad
                             J_{eff} = (\alpha+1)/4 \nonumber. 
\end{equation}
This is the identical result as obtained from direct variational 
calculations for the $S=1$ chain \cite{KenT92}.

For the $S=1-$chain the hidden order is characterized by the string
order parameter as introduced in Ref. \onlinecite{NijR87}. In our
approach, the AKLT value $g(\infty)=\frac{4}{9}$ is reproduced in the
limit of zero singlet weight, i.e.\ for $B=0$. As is seen from the
monotonic behaviour of the singlet weight in fig.~\ref{figsing}(c) in
combination with eq.(\ref{string}), the string order parameter
$g(\infty)$ decreases monotonically when $\gamma$ is increased from $-
\infty$ towards $0$.

\section{Numerical Results}

We have computed the ground state energy per rung and the 
energy gap for different values of 
the nnn interaction $\alpha$ and the alternating bond exchange
$\gamma$ in order to compare these exact results to those 
obtained in the variational calculation of the
last section in the various regions of the phase diagram.
The paths which we have investigated numerically
are shown in fig.~\ref{phasediag} as dashed lines. The results for 
points in the upper half plane follow from the symmetry property 
given in eq.(\ref{transf}). 

Numerical calculations were done on the MPP CRAY T3D SC256 of the 
Zuse Computing Center Berlin.
We used the Lanczos technique to determine the ground state energy
{\em per rung} as well as the singlet-triplet energy gap on $ 2 \times L $
lattices for $L=6,8,10,12,14$. In order to extrapolate the results 
for finite $L$ to the bulk limit we used the ansatz of 
ref.\onlinecite{BarDRS93},

\begin{equation}
        f(L) = f( \infty) + c_0 e^{-L/L_0} \: L^{-p},
                                       \label{fit}
\end{equation}

i.e.\ a power law in $L$ multiplied by an exponential.
Following ref.\onlinecite{BarDRS93} the Lanczos data were fitted
using $p=2$ for the ground state energy per rung and $p=1$ for the 
gap energy. The typical relative mean square error of these fits 
was about $10^{-10}$ for the ground state energy and about $10^{-7}$ 
for the gap energy. The fit parameter $L_0$ reflects the 
characteristic length of the system and corresponds to the 
correlation length in the chain under consideration. 
We have checked the accuracy of our program by recalculating data for
the isotropic ladder ($\alpha =2,\gamma=-1$) and have fully reproduced
the results of ref.\onlinecite{BarDRS93}.

In our numerical calculations we have concentrated on two different 
aspects of generalized spin ladders: 

\begin{itemize}
\item On the variation of the ground state energy and of the 
excitation gap when we
move from the dimer state to the Haldane limit. These data allow 
to estimate quantitatively the accuracy of the variational 
approach based on MP states and to discuss the issue whether these 
two limiting states are separated by a phase boundary.

\item On the neighbourhood of the gapless line (a) connecting the
two conformal points $\gamma=0, \: \alpha=0$ (HAFM) and 
$\gamma=0, \: \alpha = \alpha_{cri} \approx 0.5$. Such data are expected 
to contribute to an understanding of the crossover from the gapless 
phase on the critical line $\gamma = 0, \: \alpha < \alpha_{cr}$ to 
the gapped phase(s) of the MG/dimer-- and Haldane--type.
\end{itemize}

We present the variations of the ground state energy, the gap energy 
and the length $L_0$ from the dimer point to the Haldane limit on the
paths shown as dashed lines in fig.~\ref{phasediag}. Our numerical   
results for various chain lengths and their extrapolation to the bulk 
limit are given in tables~\ref{tone} and \ref{ttwo} and displayed in 
figs.~\ref{symmetryline} and \ref{figsing}. An extrapolation to the limit 
$\gamma \rightarrow - \infty$ was performed with a fourth order
polynomial fit to $E_0/L - (1+\gamma)/ 4$ for the points 
$\alpha=2, \; \; \gamma=-2.2,-2.6,-3.2,-4.4,-13$. This extrapolation 
resulted in 

\begin{equation} 
\frac{1}{J_{eff}} \left( \frac{E_0}{L}- \frac{1+\gamma}{4} \right) 
\approx -1.4069, \qquad 
\frac{1}{J_{eff}} \: \Delta \approx 0.4106, \nonumber 
\end{equation}
with $J_{eff}= (\alpha+1)/4 = 3/4$. Both numbers are in excellent
agreement with the well-known results for the ground state energy and
the gap of the Haldane chain \cite{GolJL94}. The analogous
extrapolation for the characteristic length $L_0$ (see eq.(\ref{fit})) is
shown in fig.~\ref{figsing}(d); $L_0$ increases monotonically when we
approach the Haldane limit on our path. The variation of $L_0$
corresponds to the variation of the correlation length $\xi$ although
the limiting value of $L_0$ is found somewhat larger than the accepted
numerical value for the correlation length of the Haldane chain $\xi
\approx 6.2$.

The numerical results are in agreement with the observation already
made for the variational results: On the continuous path through the
phase diagram the quantities $E_0, \: \Delta, \: L_0$ vary smoothly
and there is no indication of a phase boundary in the variation  
of these quantities. These data therefore support the hypothesis
that the dimer point, the Majumdar-Ghosh point and the Haldane
limit all are in the same phase. The gap characterizing 
these three limiting chains as well as the intermediate systems,
including the regular spin ladder thus appears only quantitatively 
different and these systems therefore should be considered 
as manifestation of basically the same quantum condensation 
phenomenon.


We now turn to the second region of interest in the phase diagram,
the area around the line of vanishing gap connecting the two known
conformal points.
We calculated the energies of the ground state and of the first 
excited states

\begin{itemize}
\item for $0.3 < \alpha < 0.6$ and small values of $\gamma$, allowing 
an extrapolation to $\gamma = 0$ 

\item for values on the line $\gamma = - \alpha$, i.e.\ on a path from the 
gapless HAF towards the gapped phase at larger values of $\alpha, \:\gamma$.   
\end{itemize}

In fig.~\ref{gap} and in table ~\ref{tthree} we show the behaviour of
the gap when the gapless line, $\gamma = 0$, is approached, i.e.\ the
vanishing of the gap on the critical line and the (small) finite gap
energy for $\alpha > \alpha_{cr}$. For $\vert \gamma \vert \ll 1$ and
two values of $\alpha < \alpha_{cr}$ we have analyzed the results
assuming a power law behaviour, $\Delta \propto \vert \delta \vert^y $,
(where $\delta = \gamma / (2+\gamma)$ is the dimerization parameter) in
order to compare with the prediction $y=2/3$ by Cross and Fisher
\cite{CroF79}. From a log-log plot (see insert in fig.~\ref{gap}) we find
$y \approx 0.862 $ for $\alpha = 0.30$ and $y \approx 0.795 $ for 
$\alpha = 0.44$, i.e.\ an $\alpha$-dependent exponent which is somewhat larger 
than predicted in ref. \onlinecite{CroF79}.

The gap close to the conformal 
line is found at wavevector $k=0$, whereas the gap in the Haldane 
phase is known to occur at wavevector $k= \pi$ (the wavevector $k$ is
defined through the factor $\exp(\imath k)$ which multiplies the 
wavefunction upon the transformation $j \to j+1$, i.e.\ a shift of 
rungs by one unit). To discuss this crossover in the wavevector
of the lowest excitation we have calculated the lowest excitation 
energies for these two wavevectors, $\Delta_{k=0}, \; \Delta_{k=\pi}$ 
along the path $\gamma =- \alpha$. The results of an extrapolation to 
the thermodynamic limit as described above are shown in fig.~\ref{gap2}. 
It is seen that the quantity $\Delta_{k=0} - \Delta_{k=\pi}$ changes 
sign for $\alpha \approx 0.55$ i.e.\ close to the line
$\gamma = -1 + \alpha$ which connects the dimer and MG points.
These results for the gap (and the corresponding results for the 
ground state energy) join smoothly to the results presented before 
when values $\alpha = 2$ are reached. Actually the excitation energies
on the disorderline $\gamma = -1 + \alpha$ can be calculated in 
perturbation theory in $\alpha$ to a high degree of accuracy for 
$\alpha \ll 1$. For $k = \pi$ we have the exact result $\Delta_{k=\pi} = 1$ 
on the whole disorder line, as will be published elsewhere. 

Generally, our numerical results confirm the variational results in
that variations over most of the phase diagram are smooth and do not
indicate any phase boundaries. However, we find that the excitation
spectrum does change qualitatively along a path from the critical line
to the gapped phase; this may be another aspect of the fact that the
maximum of the structure factor $S(k)$ is found at finite $k$ and may
be related to the existence of a spiral phase as speculated before
\cite{ZenP95,ChiPKSR95}. In order to clarify this point, more
detailed results on the excitation spectrum in this region of the
phase diagram should be obtained. However, since a reliable finite
size analysis is not possible for general values of the wavevector we
cannot discuss the character of the complete excitation spectrum with
comparable accuracy and therefore cannot decide whether the minimum
excitation energy occurs at some finite value of the wavevector.
In order to provide some first information we present in
fig.~\ref{spectrum} the excitation spectrum for the finite ladder with
$2 \times 12$ sites at the points $\alpha = -\gamma = 0.54, \: 0.56, \:
0.58$. The minimum excitation energy is found at $k=0$ for $\alpha =
-\gamma = 0.54$, at $k = \pi$ for $\alpha = -\gamma = 0.56$. 
For $\alpha = -\gamma = 0.56$ the minimum is found at finite wavevector, 
$k \approx 2 \pi/3$, 
although dispersion is very weak and it is not clear whether the
minimum at finite wavevector will survive an appropriate extrapolation
to the infinite system. Actually, this result may be more accurate
than one might expect from the small size of the system since we found
in the finite size extrapolations for $k=0,\pi$ that the results for 
the finite $2 \times 12$ site system differ very little (relative difference
$10^{-6}$) from the thermodynamic limit. For a full understanding more
calculations are clearly required.

\section{Conclusions}

We have investigated the phase diagram of a generalized spin ladder
with additional interaction on diagonal bonds. This model includes
several known cases, in particular the $S=1$ chain in the Haldane
phase and the dimer and Majumdar-Ghosh chains belonging to a dimerized
phase.  This generalized spin ladder was studied with a variational
wave ansatz and numerical techniques. Using the concept of hidden
order throughout the whole phase diagram (with the exception of the
line of gapless points), a rotationally invariant matrix-product state
was proposed for the ground state of these ladders.  The ground state
energy, and spin and string correlations were calculated. The
comparison with numerical results for the ground state energies leads
to agreement typically within about 4 - 5\%, consistent with the 
qualitative nature of our
approximation. In the $S=1$ chain limit the well known AKLT state and 
the variational energy of 4/3 per site are reproduced and extrapolation
of the numerical results leads to excellent agreement with the known
values for the ground state energy and the gap of the Haldane chain. A
string correlation function was chosen such as to reproduce in the
corresponding limit the AKLT value of 4/9 for the factorizing
approximation to the $S=1$ chain. This string correlation function is
directly related to the singlet weight in the MP wave function and
decreases monotonically when one moves from the Haldane limit towards 
dimerized states.

>From both the numerical and the variational calculations we find   
on paths connecting
the dimer and Majumdar-Ghosh points to the Haldane limit
smooth variations of all quantities considered. We are 
therefore lead to conclude that these dimerized af $S=\frac{1}{2}$ chains  
and the af $S=1$ chain are in the same phase and that the gap in the
excitation spectrum of these chains is of the same nature.   
On the other hand, our numerical data for the excitation spectrum 
close to the line connecting the dimer and the MG points show 
a shift of the wavevector of the minimum excitation energy from 
$k=0$ to $k=\pi$ as well as an indication that the minimum may be 
found at intermediate wavevectors. Thus the question of the existence
of a spiral phase remains unresolved and requires further investigation. 

Using the present approach elementary excitations can be
constructed as soliton-like states as for dimer-like systems
\cite{ShaS81} and for the $S=1$ chain \cite{NeuM96}. Our approach
can also be extended to include anisotropic couplings; in particular,
similar as in the af $S=1$ chain, there
should be a critical value for the Ising-like anisotropy on the legs,
at which the system exhibits long range order even for 
$\gamma \to - \infty$. These further applications of the MP approach
are now under investigation.

\section*{acknowledgements} 
We gratefully acknowledge useful discussions with A. Kolezhuk,
B. L\"uthi and C. Waldtmann. This work was supported by the German
Federal Minister of Research and Technology (BMBF) under contract
number 03-MI4HAN-8. The numerical calculations were performed at the
Regionales Rechenzentrum Niedersachsen and Zuse Rechenzentrum Berlin,
we wish to thank these institutions for their helpful cooperation.

\newpage
\begin{table}
\caption{Ground state energy $E_0$ and gap energy $\Delta_{k=\pi}$ for
$\gamma = -1$ and various values of $\alpha$.
\label{tone}}
\end{table}

\begin{table}
\caption{Ground state energy $E_0$ and gap energy $\Delta_{k=\pi}$ for
$\alpha = 2$ and various values of $\gamma$.
\label{ttwo}}
\end{table}

\begin{table}
\caption{Gap energy $\Delta_{k=0}$ close to the gapless line (a). 
\label{tthree}}
\end{table}

\newpage

\begin{figure}
\caption{(a) Ground state energies, (b) correlation lengths and (c) singlet
        weight on the lines $\gamma = 0$ and $\gamma = -1$.  
        Variational approach (full lines) vs. numerical results ($\Diamond$). 
        \label{symmetryline}}
\end{figure}

\begin{figure}
\caption{Approach to the Haldane limit on the line $\alpha = 2$: 
        (a) ground state energy, (b) gap energy, (c) singlet weight,
        (d) characteristic length $L_0$.
\label{figsing}}
\end{figure}

\begin{figure}
\caption{$\gamma$-dependence of the energy gap $\Delta_{k=0}$ close to 
        the gapless 
        line (a): $\alpha = 0.30 (+),\: 0.44 (\times),\: 0.52 (\Diamond),
        \: 0.60 (\Box)$. 
        Insert: Scaling of the gap with the dimerization parameter
        $\delta = \gamma / (2+\gamma)$ for $\alpha = 0.30, \: 0.44$.
\label{gap}}
\end{figure}

\begin{figure}
\caption{Energy gaps $\Delta_{k=0}$ and $\Delta_{k=\pi}$ along the line 
        $\gamma = -\alpha. \: \: \delta = (\alpha^2 + \gamma^2)^{1/2}.$  
        Full lines are guides to the eye.
\label{gap2}}
\end{figure}

\begin{figure} 
\caption{Excitation spectra $\epsilon(k)$ close to the disorder 
      line, $-\gamma = \alpha = 0.54,\: 0.56,\: 0.58$ for $L=12$. 
      Insert: Expanded scale for $-\gamma = \alpha = 0.56$.
      Full lines are guides to the eye.
\label{spectrum}} 
\end{figure}

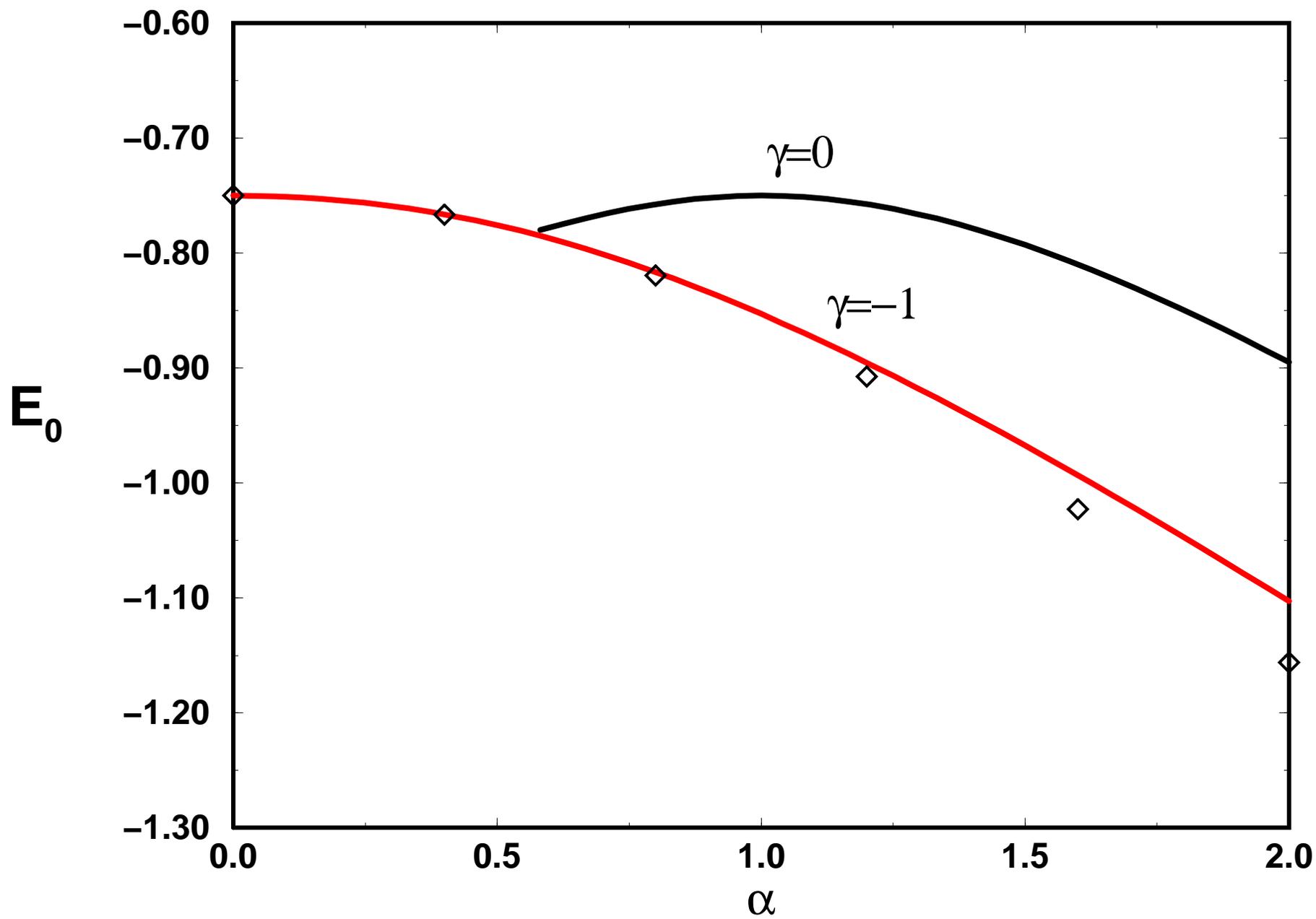
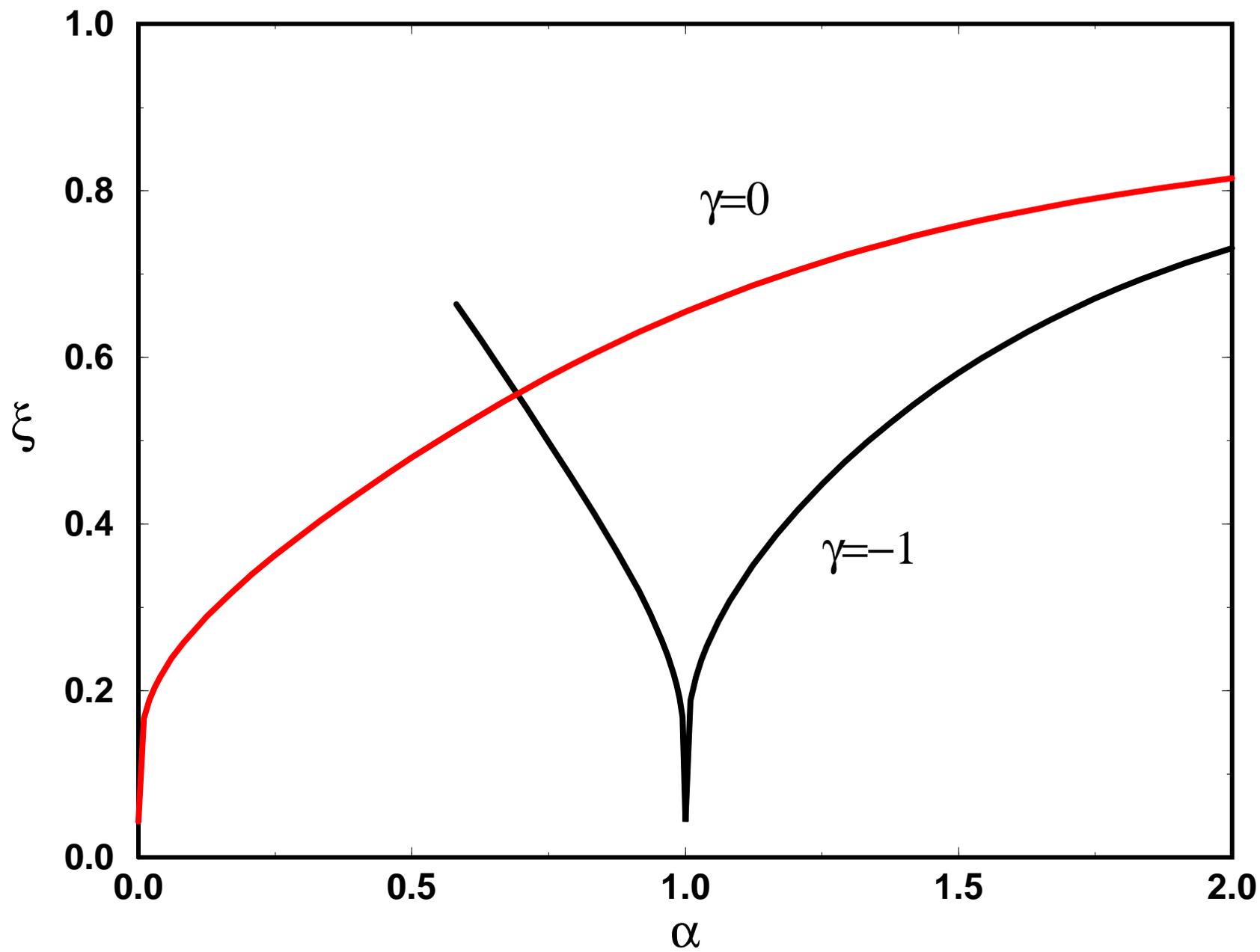
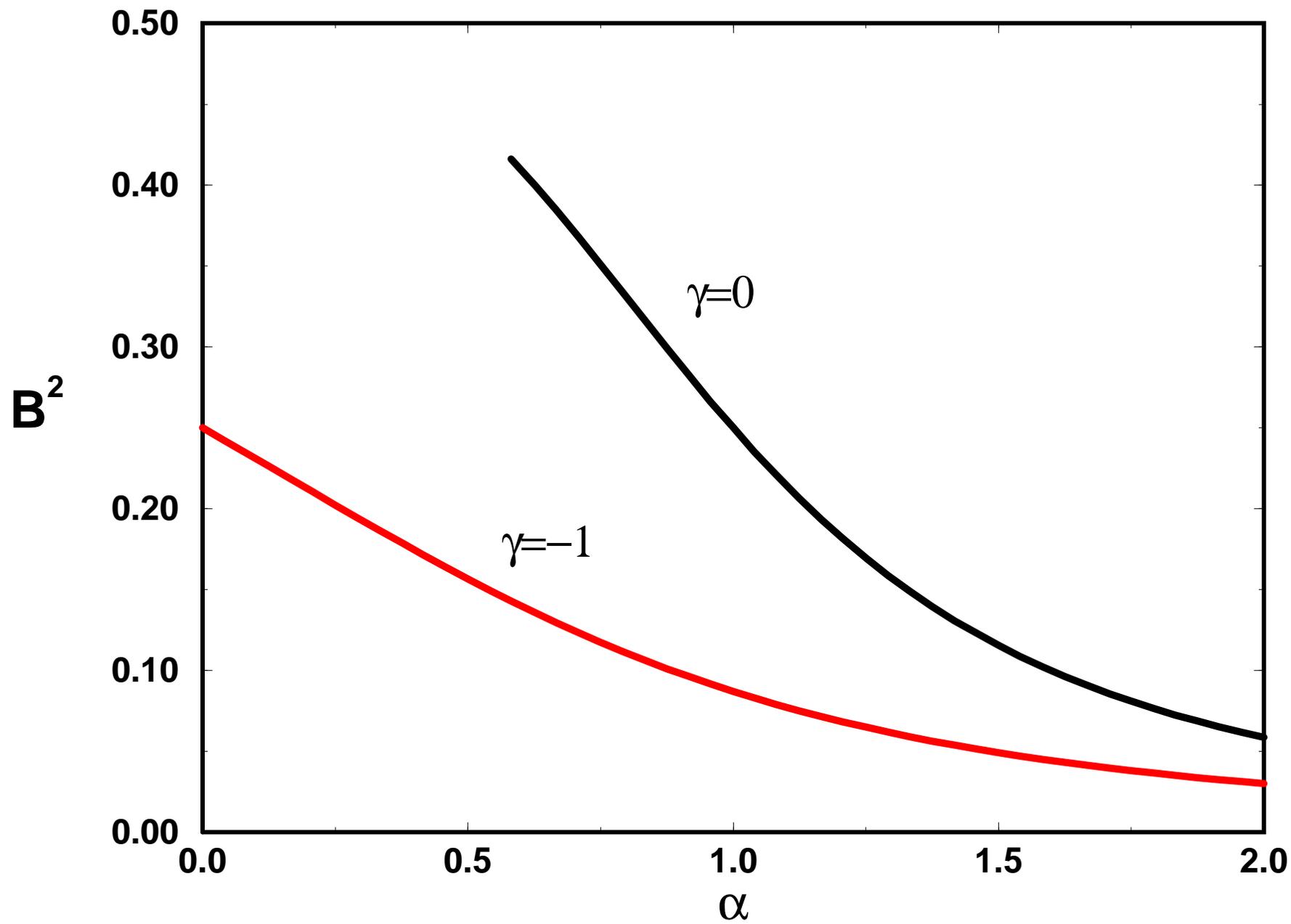
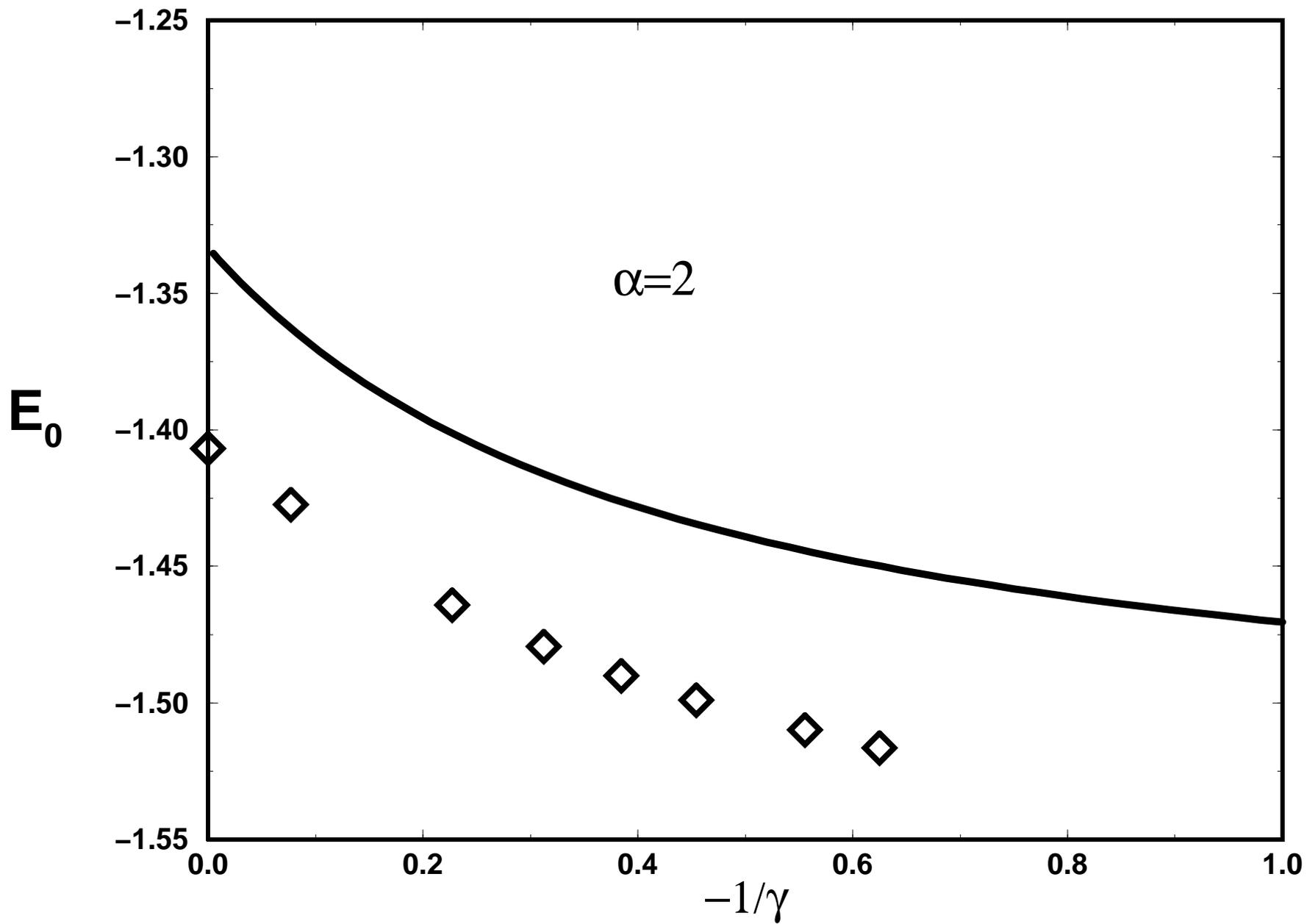
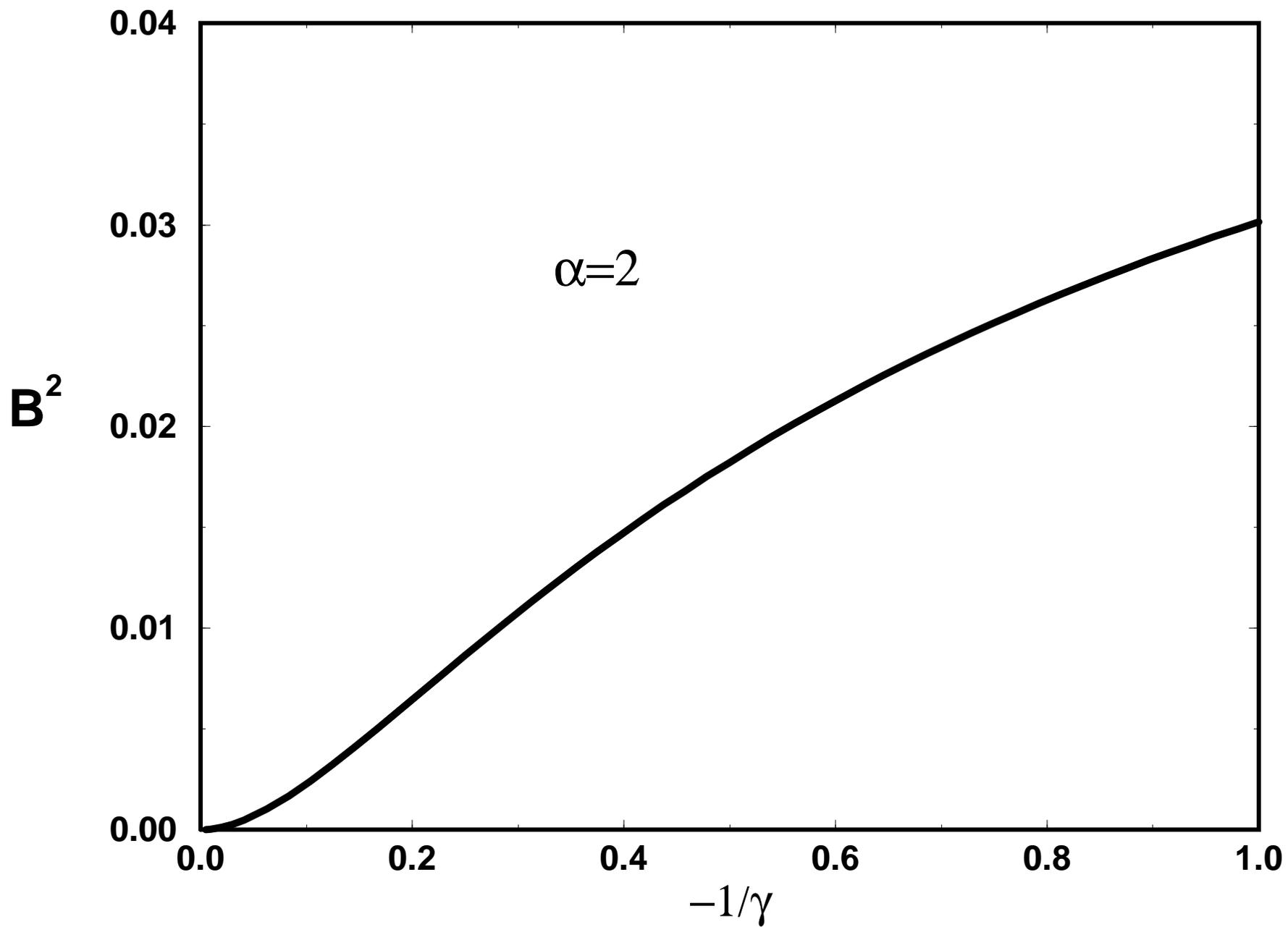
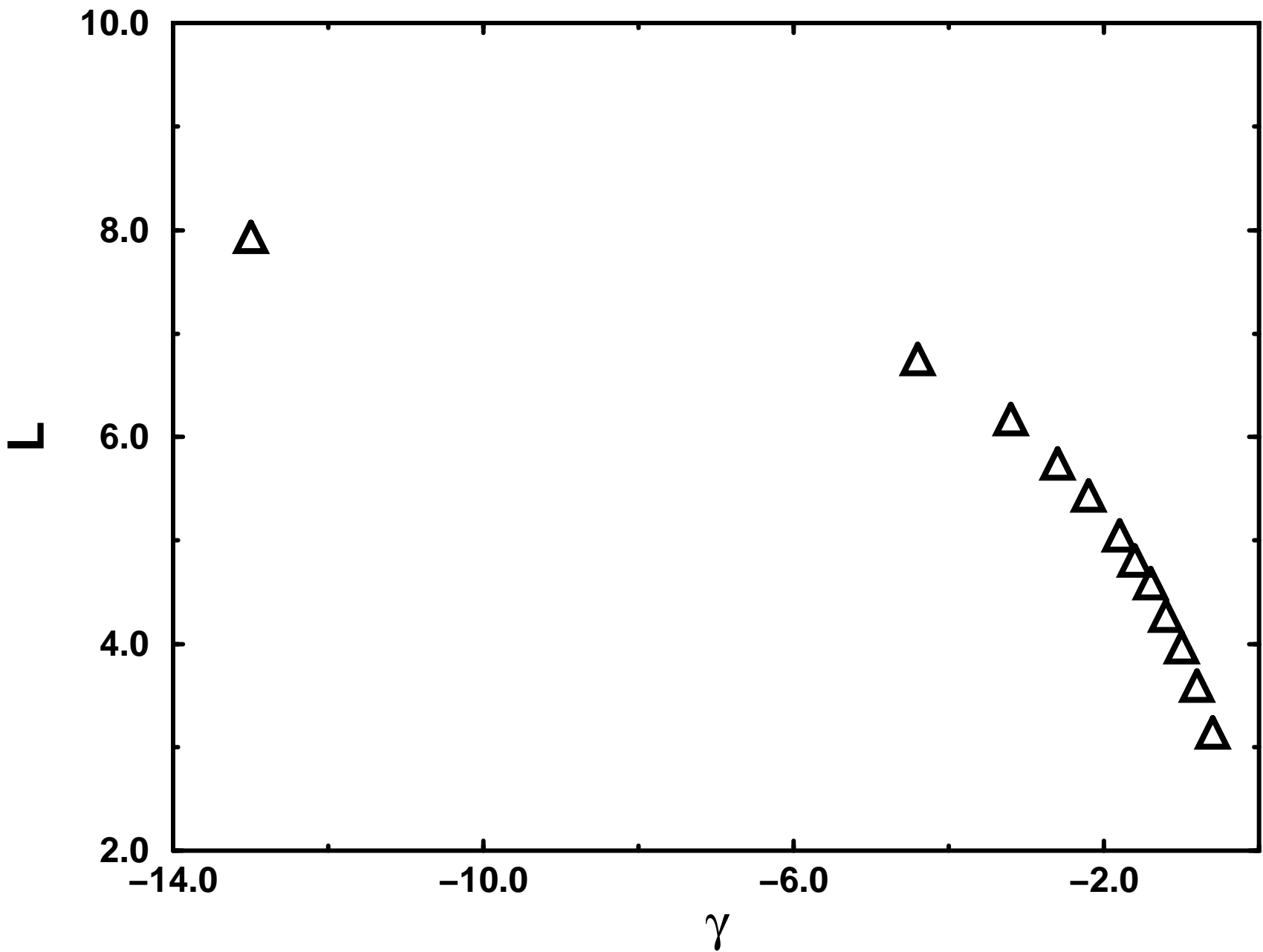
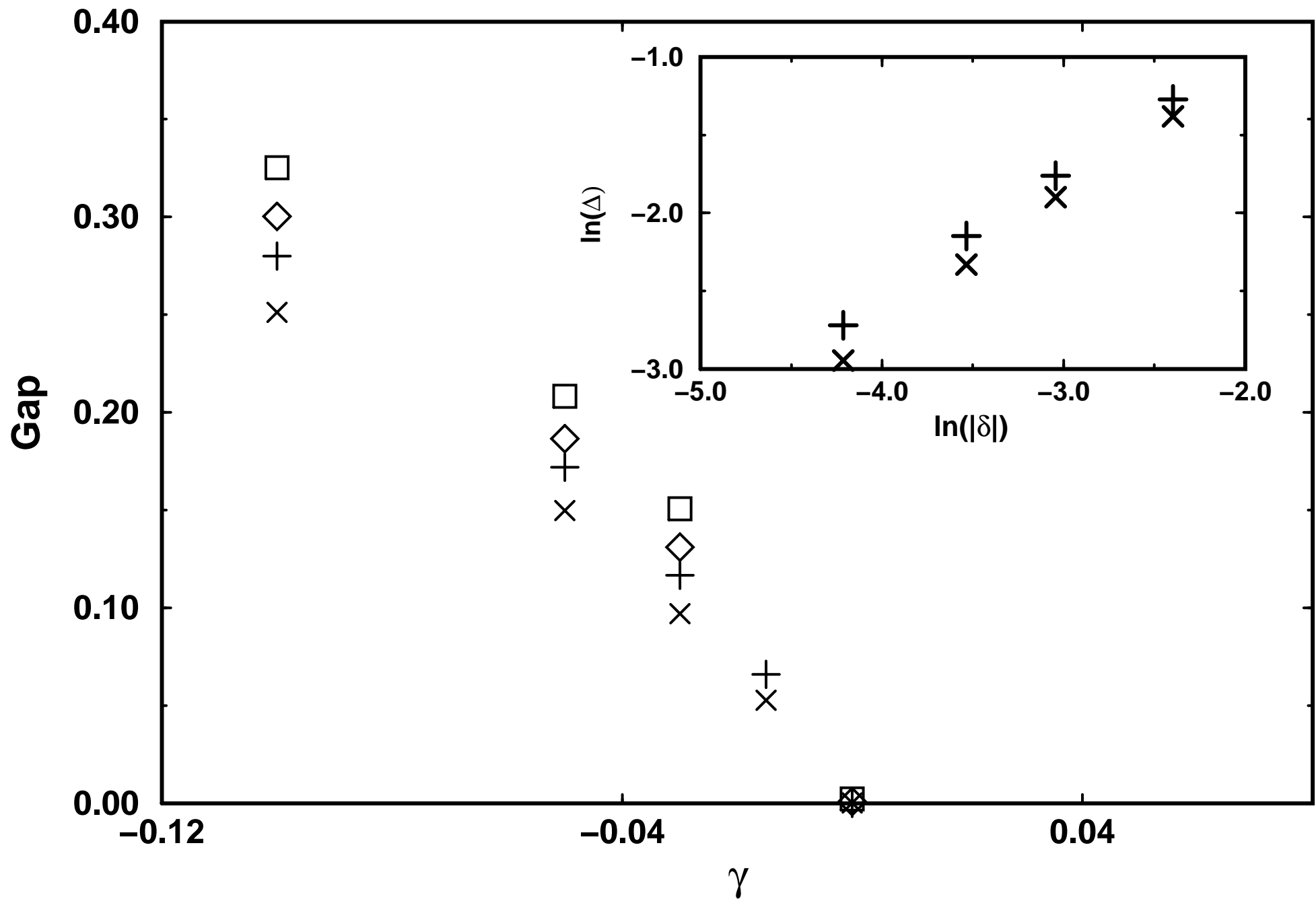
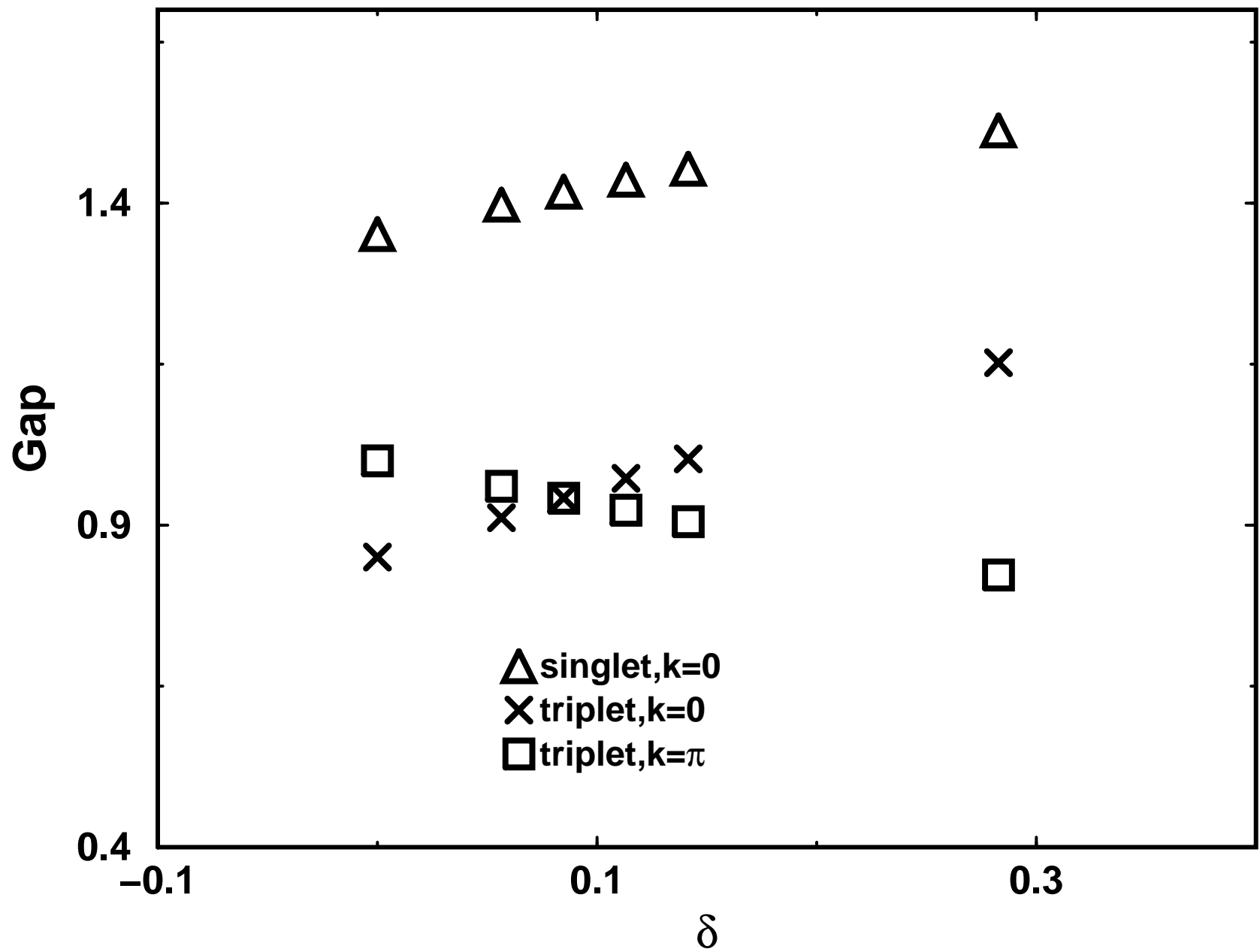
\end{document}